\documentclass{article}

\usepackage{arxiv}

\usepackage[utf8]{inputenc} 
\usepackage[T1]{fontenc}    
\usepackage{hyperref}       
\usepackage{url}            
\usepackage{booktabs}       
\usepackage{amsfonts}       
\usepackage{nicefrac}       
\usepackage{microtype}      
\usepackage{lipsum}
\usepackage{graphicx}
\usepackage{float}
\graphicspath{ {./images/} }

\title{AI and personalized learning: bridging the gap with modern educational goals}

\author{
 Kristjan-Julius Laak \\
  Institute of Computer Science\\
  University of Tartu\\
  Tartu, Estonia \\
  \texttt{julius.laak@gmail.com} \\
   \And
 Jaan Aru \\
  Institute of Computer Science\\
  University of Tartu\\
  Tartu, Estonia \\
  \texttt{jaan.aru@gmail.com} \\
}

\begin{document}
\maketitle
\begin{abstract}
Personalized learning (PL) aspires to provide an alternative to the one-size-fits-all approach in education. Technology-based PL solutions have shown notable effectiveness in enhancing learning performance. However, their alignment with the broader goals of modern education is inconsistent across technologies and research areas. In this paper, we examine the characteristics of AI-driven PL solutions in light of the goals outlined in the OECD Learning Compass 2030. Our analysis indicates a gap between the objectives of modern education and the technological approach to PL. We identify areas where the AI-based PL solutions could embrace essential elements of contemporary education, such as fostering learner’s agency, cognitive engagement, and general competencies. While the PL solutions that narrowly focus on domain-specific knowledge acquisition are instrumental in aiding learning processes, the PL envisioned by educational experts extends beyond simple technological tools and requires a holistic change in the educational system. Finally, we explore the potential of generative AI, such as ChatGPT, and propose a hybrid model that blends artificial intelligence with a collaborative, teacher-facilitated approach to personalized learning.
\end{abstract}


\section{Introduction}

Prominent educational researchers and organizations agree that education must be learning-centered for every pupil to reach their potential (Bransford et al., 2000; Groff, 2017; O’Brien et al., 2009; OECD, 2006; Vygotsky \& Cole, 1978). The learning-centered paradigm stems from changes in industry and society: The predominantly knowledge-based form of work requires an education system that values individual differences and promotes lifelong learning skills (Lee, 2014). Since early experiments demonstrated the benefits of one-on-one tutoring over group-level instruction (Bloom, 1984; Salvin \& Karweit, 1985), researchers have sought technological solutions to achieve cost-effective personalized tutoring (Lee et al., 2018). Personalized learning (PL) aims to achieve this goal by tailoring instruction, pace, methods, and content to the interests, needs, and goals of individual learners (Beese, 2019; US Department of Education, 2010; Walkington \& Bernacki, 2020). Breakthroughs in technology and artificial intelligence (AI) have led to a rapid increase in the number of technological solutions to PL (Chen et al., 2020). Over the last decades, AI-driven adaptive learning systems (Corbett et al., 1997; Maghsudi et al., 2021; VanLehn, 2011) have been researched and developed to provide learners with individualized lesson sequences, content recommendations, tasks, and automated assessments (Zawacki-Richter et al., 2019). A prevailing view in the AI and Education (AIEd) literature is that personalized adaptive learning systems provide equitable access to high-quality education, contrasted with a “traditional,” one-size-fits-all approach (Tetzlaff et al., 2020; Maghsudi et al., 2021; St-Hilaire et al., 2022). However, previous reviews have pointed to limited analysis of the educational underpinnings of AIEd (Chen et al., 2020; Hinojo-Lucena et al., 2019; Zawacki-Richter et al., 2019).

The belief in technology-based personalized learning systems is not new, originating from the 20th-century behaviorists’ theories and has maintained its appeal despite the evolving societal needs and educational expectations (Watters, 2023). One framework for personalized learning was outlined in the early 1960s in Richard D. Smallwood’s dissertation on “teachable machines” (Smallwood, 1962). As envisioned by Smallwood, these machines would provide an individualized pace of learning, apply repetition until mastery, provide immediate feedback, and record student performance (Smallwood, 1962, pp. 2–3 as cited by Essa 2016). The need for individualized learning was fueled by experiments that demonstrated higher learning gains in 1:1 tutoring than in traditional group instruction, named the 2-sigma problem (Bloom, 1984; Salvin \& Karweit, 1985). Today, Smallwood’s vision is reflected in the widely used definition of the U.S. Department of Education (2010, 2017), which refers to PL as “instruction in which the pace of learning and the instructional approach are optimized for the needs of each learner.” This narrow interpretation of PL as an optimization problem has become prevalent in the AIEd literature, leading to the conflation of “personalized learning” with “adaptive learning systems” (e.g., Bernacki et al., 2021; Maghsudi et al., 2021; Shemshack et al., 2021; Shemshack \& Spector, 2020; Tetzlaff et al., 2021). The technological approach to PL diverges from the more holistic PL paradigm that has advocated for a broader change in the educational system, including but not limited to educational strategies, curricula, and school managements, in which technology has only a supportive role (Hopkins, 2010; Lee, 2014; Lee et al., 2018; Miliband, 2006). However, AIEd is still in its infancy, presenting an opportunity to shape its development in alignment with the needs of modern education.

Here, we analyze this technological approach to PL through the lens of the educational goals outlined in the OECD Learning Compass 2030, using it as a framework to support our arguments (OECD, 2019b). In addition to core knowledge and skills, the framework highlights three main goals of education: focus on general competencies, developing learners’ agency, and building on the Anticipation-Action-Reflection (AAR) cycle. First, the AAR cycle reflects the need for deeper cognitive and metacognitive engagement before, during and after learning (Bransford et al., 2000; Chi \& Wylie, 2014; Dehaene, 2020). This engagement is largely mediated by self-regulated learning (SRL) skills that are reflected in the ability of the learner to plan their learning activities, set and take action towards goals, monitor their level of understanding during the learning, and evaluate their performance afterwards (Knowles, 1975; Zimmerman, 1986, 2000; Saks \& Leijen, 2014; Panadero, 2017). For example, it has been consistently shown that without proper SRL skills, learners overwhelmingly choose the least effective strategies for learning (McCabe, 2011). The second goal - learners’ agency - is a broad and interdisciplinary term tightly connected to SRL that denotes the skills that enable the learner to take initiative in leading their learning throughout life (Brod et al., 2023; Council of the EU, 2002; OECD, 2006). Third, we define general competencies as personal attributes or behaviors developed through action, experience, and reflection (UNESCO, 2017). Examples of these competencies include critical thinking, collaboration skills, and complex problem-solving skills (OECD, 2019b; UNESCO, 2022). 

Our main research question was whether the technological approach to PL aligns with the goals of modern education as outlined in the Learning Compass 2030. We explore this question by examining how the literature on adaptive learning systems emphasizes certain aspects of modern education while neglecting others. The articles were gathered by searching Web of Science database using various search terms (i.e., “personalized learning”, “adaptive learning system”, “intelligent tutoring system”, “AIEd”, “AI and education”). We included both empirical studies and reviews about the use of adaptive learning systems in K-12 and higher education written in English and published in peer-reviewed journals.

Based on the included articles, we first bring out the benefits of adaptive learning systems. Next, we explore key issues with the technological paradigm of personalized learning. In the final section, we explore the potential of generative AI and propose a hybrid framework that envisions a hybrid relationship between teachers, GenAI, and learners.

\section{Benefits of adaptive learning systems}
The benefits of adaptive learning systems are manifold, including flexibility in time and location, timely feedback, and faster student progression (Moreno-Guerrero et al., 2020; Pliakos et al., 2019), offering a more dynamic approach to knowledge acquisition compared to group-level instruction (Luckin et al., 2016). These advantages stem from the typical three core components of adaptive learning systems: a content model, a student model, and a pedagogical model (Martin et al., 2020). The pedagogical model identifies and sequences learning objectives, guides learners through individual learning paths, and provides timely feedback and assessment (Ezzaim et al., 2022; Martin et al., 2020; Pelánek, 2017). The pedagogical model usually provides distributed practice until the learning objective is mastered, an effective learning strategy for long-lasting comprehension and memory formation (Dempster \& Farris, 1990; Gerbier \& Toppino, 2015; Hintzman, 1974; Soderstrom \& Bjork, 2015; Srinivasan, 2022; Tetzlaff et al., 2021). The pedagogical model provides continuous corrective feedback, which is a highly effective teaching strategy for empowering learners to improve their understanding (Dehaene, 2020; Hattie, 2011; Hattie \& Timperley, 2007; Osakwe et al., 2022). The student model stores learner characteristics and tracks the learner's evolving level of understanding (Greer \& McCalla, 1991). Together with the pedagogical model, dynamic student models allow task difficulty to be adapted to the individual’s zone of proximal development, where tasks are neither too boring nor too difficult to cause discouragement, making learning naturally engaging and challenging (Bransford et al., 2000; Vygotsky \& Cole, 1978). The individual learning steps are recommendations on the content model, a digitized, connected domain knowledge graph (Bartl \& Belohlavek, 2011). The presentation of appropriate levels of tasks is continued until the student has mastered the topic at hand (Shute \& Towle, 2010). In short, PL systems that monitor each student’s progress and adapt learning content, methods, and assessment to individuals on an ongoing basis (Shemshack \& Spector, 2020; Xie et al., 2019; Vandewaetere \& Clarebout, 2013) are efficient for mastering certain domain knowledge and improving learning performance (Walkington \& Bernacki, 2020; Cui et al., 2018; Feng et al., 2018; Steenbergen-Hu \& Cooper, 2014; Urh et al., 2015). Hence, we recognize the value of adaptive learning systems as supportive tools for knowledge acquisition.

In addition, a substantial body of research has been done to design specific digital PL systems that try to promote student SRL skills, enhance learner’s agency, and engage students cognitively (e.g., Azevedo et al., 2022; Biswas et al., 2016; du Boulay, 2019; Long \& Aleven, 2017). Comprehensive theoretical and empirical basis for designing learning environments that foster SRL and metacognition have been provided (for an overview, see Azevedo \& Aleven, 2013). Scaffolding SRL behaviors has been an important research focus in the field of intelligent tutoring systems (ITS). For instance, much work has been done with an ITS called MetaTutor that scaffolds SRL processes, such as goal setting, evaluation of learning strategies, self-monitoring of progress, and supports cognitive processes of learning required for long-term memory retention, e.g., note-taking, connecting prior and new knowledge, hypothesis generation (Azevedo et al., 2022). Extending domain-level ITS systems with metacognitive tutoring has positively affected student help-seeking strategies and learning outcomes (Aleven et al., 2006; Hacker et al., 2009; Roll et al., 2007). For example, extensive studies have been done with a cognitive science tutoring system called Betty’s Brain, where middle-school students learn by teaching a virtual agent and creating causal models (Biswas et al., 2005). Leelawong \& Biswas (2008) showed that Betty’s Brain led to higher performance and skill transfer to novel domains with self-regulation features such as self-monitoring, goal setting, and self-assessment. In Betty’s Brain, students must also develop and apply metacognitive strategies for goal setting, planning, and evaluation to make effective decisions from information-seeking to solution assessment (Biswas et al., 2016). While some of the aspects of the Learning Compass have been incorporated into dialogue-based ITS (such as Betty’s Brain and Metatutor), the majority of adaptive learning systems do not usually employ these aspects (Ezzaim et al., 2022).  Many distinguished researchers agree that adaptive learning systems are narrowly and primarily focusing only on domain-knowledge delivery (e.g., Brod et al., 2023; Järvelä et al., 2021; Molenaar, 2022a). Next, we explore some specific issues with adaptive learning systems in light of the goals of modern education.

\section{Challenges of personalized learning systems}
\subsection{Focus on performance}
Adaptive learning systems are predominantly subject-based platforms that dynamically adapt instruction to individual learners’ characteristics, such as learning style and ability, to maximize learning achievement (Capuano \& Caballé, 2020; Kerr, 2016). While recognizing the extensive research done by learning scientists behind some ITS mentioned above (Azevedo \& Aleven, 2013; Azevedo et al., 2022; Aleven et al., 2006; Hacker et al., 2009; Roll et al., 2007; Biswas et al., 2005; Leelawong \& Biswas, 2008; Biswas et al., 2016; du Boulay, 2019; Munshi et al., 2023), the adaptive learning literature is overwhelmingly focused on learning achievements, and only a fraction measure other cognitive aspects such as higher-order thinking and collaboration/communication (Xie et al., 2019). In other words, the main goal of most PL technologies is to increase learning efficiency (Li \& Wong, 2021). The efficiency of adaptive learning systems is measured by the increase in learning gains in a given time, i.e., the difference between pre- and post-test scores. Studies of adaptive learning platforms show their efficiency over non-adaptive approaches when measured by student scores and learning speed (Ma et al., 2014; Steenbergen-Hu \& Cooper, 2014; Roschelle et al., 2016). For example, students instructed by Yixue (Squirrel AI) scored up to 456\% higher in less time than students in traditional classrooms (Cui et al., 2018; Feng et al., 2018; H. Li et al., 2018). The AI-powered PL platform Korbit showed 2.5 times higher scores compared to a non-adaptive Moodle course (St-Hilaire et al., 2022). The Rimac physics tutoring system has been shown to produce better learning outcomes for high and low prior knowledge students when the system has a dynamic student model compared to a non-adaptive one (Katz et al., 2021, p. 202). A meta-analysis of post-test scores found that ITS for medical practice and physics had the same effect size as 1:1 human tutors (VanLehn, 2011). While the efficient acquisition of foundational knowledge is certainly commendable, it is essential to recognize that modern education aims to achieve more than the efficiency of getting higher test scores. Conversely, a focus on performance is often associated with traditional, teaching-centered classrooms, where factors such as learner self-regulation, agency, well-being, and conceptual understanding may not be given the same level of priority. Notably, there is a major concern with the exclusive focus on learning performance.

A century of experimentation has shown that it is critical to distinguish learning - the long-term ability to exhibit new behaviors and knowledge - from performance - temporary fluctuations in knowledge and behavior measured during acquisition (Metcalfe et al., 1994). An integrative review illustrates why performance is a highly imperfect index of long-term learning (Soderstrom \& Bjork, 2015). First, the authors provide extensive evidence that learning can occur without changes in performance. For example, according to a well-documented psychological phenomenon called “latent learning,” learning often becomes visible only after a reward or incentive is provided (Tolman \& Honzik, 1930). These early results have been confirmed in human studies and suggest that latent learning, i.e., learning that does not improve performance, may be more effective than learning in which rewards (e.g., scores) are given at the beginning of the learning experience. Second, Sonderstrom and Bjork show how changes in performance may not correspond to changes in understanding. For example, repeating the same content over and over (massing) improves short-term performance but does not improve long-term learning. Lastly, the authors show that errors and mistakes are often interpreted as a decrease in performance but actually improve long-term retention and transfer (Soderstrom \& Bjork, 2015). In summary, measures of performance during acquisition may not be indicative of lasting learning, and measures that show degraded performance during learning may improve long-term retention and transfer (Metcalfe et al., 1994). Hence, the remarkable gain in performance and efficiency may not be an indication of a more effective approach to learning.

\subsection{Domain-specific knowledge}
A comprehensive review of empirical AIEd studies conducted between 1993 and 2020 found that the majority of educational AI systems focus only on a few subject areas: computer science and engineering (35\% of studies), mathematics (20\%), and foreign languages (10\%) (Zhang \& Aslan, 2021; see Ezzaim et al., (2022) for specific software tools by subject). Although some ITS have enabled learning-by-doing in STEM subjects and provided students with elaborate problem-solving opportunities (e.g., Koedinger et al., 2015), even these innovative solutions remain subject-specific because of their architecture and reliance on the content model (Doignon \& Falmagne, 2012; Zhang \& Aslan, 2021). Hence, given the quantitative nature of knowledge graphs, adaptive learning systems are significantly constrained in the breadth and depth of what could be effectively learned (Essa, 2016) and can only support the effective acquisition of foundational knowledge and skills (Pedaste \& Leijen, 2018). Although the latter are essential building blocks for independent thinking, the evolving demands of the modern world also necessitate the cultivation of non-epistemic competencies (as listed in the Learning Compass 2030).

It is not apparent how these non-epistemic competencies, often labeled as 21st-century skills, fit into the model of mastery-based adaptive learning (Tuomi, 2023). From the perspective of knowledge graphs, the concept of optimizing “soft skills” and the feasibility of their assessment through error-correcting and automated assessment remains an open challenge (Shemshack \& Spector, 2020). For example, what questions would effectively gauge a student’s mastery of critical thinking and creating new value? Incorporating competencies, such as critical thinking, into knowledge spaces can be a formidable undertaking. Adaptive learning systems are inherently limited to machine-assessable knowledge and skills. To unlock their potential, learners must creatively extend their learning outside the subject domain (UNESCO, 2017). Developing creative and novel ideas requires opportunities for experimentation, exploration, and problem-solving beyond the boundaries of the majority of the existing adaptive learning systems (Stanley \& Lehman, 2015; Turner, 2014).

Technically, the purportedly “personalized” learning paths fall short of delivering the freedom and choice that the more holistic paradigm of PL aspires (see Hopkins, 2010). Precisely because being limited to the same underlying knowledge spaces, AI-recommended learning paths are merely different routes for mastering the same knowledge at an individualized order and pace. In that regard, adaptive learning systems seem to optimize the traditional educational system without providing learners much agency and customization for setting their goals outside the provided task and existing subject domain.

\subsection{Limited agency and learning skills}
Rather than blindly following the predefined sequence of learning objectives, OECD Learning Compass 2030 emphasizes the need for learners to navigate unfamiliar contexts and find direction through their own agency (OECD, 2019c). Learners with agency and proficient SRL skills can set learning goals, choose actions toward them, and adjust their strategies when needed (Jarvis, 2004; UNESCO, 2017; Winne, 2017; World Economic Forum, 2021). Numerous studies have shown the importance of self-regulation for effective learning in distant, e-learning, digital, or other self-directed learning situations (Barnard-Brak et al., 2010; Bail et al., 2008; Hofer \& Yu, 2003; Azevedo et al., 2009; Puzziferro, 2008). Hence, there is a strong consensus that supporting learners’ agency and ability to self-regulate should be the priority of adaptive learning systems (Brod et al., 2023; Järvelä et al., 2021).

As described above, some ITS are designed to foster student SRL skills, agency, and other more generic competencies besides subject-domain expertise. However, even the most sophisticated adaptive learning solutions (not the ITS mentioned above) do not support learners’ agency and self-regulation (Molenaar, 2022a). Moreover, existing SRL skills are necessary study effectively with personalized learning systems (Laak \& Aru, 2024). Recognizing this issue, one K-12 adaptive platform allows students lacking SRL training to override the platform's suggestions (Cukurova et al., 2023). Notably, a recent study of a well-studied learning platform showed a significant decline in students’ SRL skills while working with the system (Harati et al., 2021). This kind of cognitive offloading has also been shown with generative AI solutions (Fan et al., 2024; Gerlich, 2025).

Dignath and Veenman (2021) bring out four essential preconditions that enhance SRL: self-direction (decision-making freedom), cooperative learning (students work together), constructive learning principles (e.g., problems have several solutions), and focus on transfer (learning is integrated into a real-life context). Similarly, information-theoretically, a system is autonomous when (1) it is not determined by its environment and (2) it determines its own goals (Bertschinger et al., 2008). Because recommendation engines effectively control learning goals, content, sequence of tasks, etc., they potentially reduce the learner’s agency and take away the freedom required for developing healthy SRL skills. The need for agency over one’s learning can be illustrated by the classic experiment (Figure 1A) in which only the kitten that could exhibit self-directed movements developed normal vision (Held \& Hein, 1963). Similarly, if the learner is passively moved along the prescribed learning path, he or she may not acquire normal learning and self-regulation (Figure 1B). The lack of support for learner’s learning skills in some adaptive learning systems has been mentioned previously (Pedaste \& Leijen, 2018). Although a framework for the adaptive assignment of agency (Brod et al., 2023) and a prototype for the transition from AI regulation to self-regulation in adaptive learning systems (Molenaar, 2022a) has been proposed, we have yet to see the emergence of human-AI hybrid intelligence systems (Cukurova, 2024).

\begin{figure}
  \centering
    \includegraphics[width=1\linewidth]{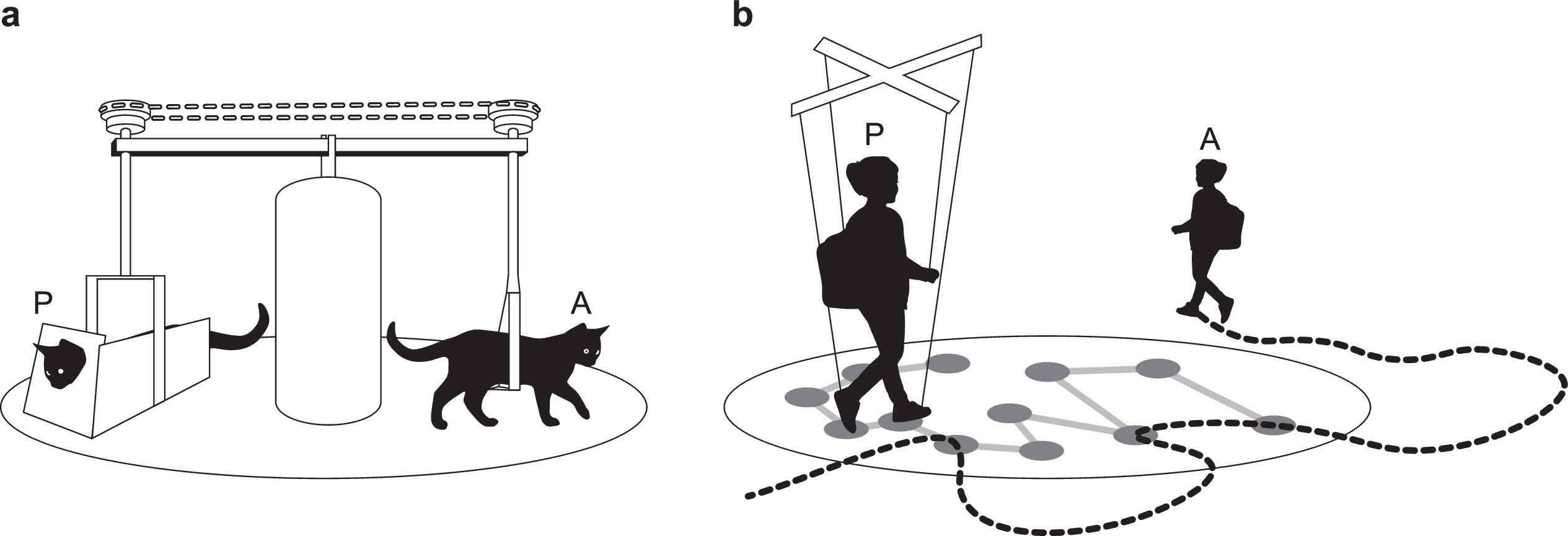}
    \caption{\textbf{a} Adapted illustration of the experimental setup of Held and Hein (1963) in which two kittens were reared in darkness from birth and learned to see the world only in a carousel. While kitten P was passively moved around in a casket, kitten A could rotate around on its own. Significantly, only kitten A, exhibiting self-directed movement, developed normal vision. \textbf{b} Illustration of two learners in an open educational field. Learner P is moved by an adaptive learning system (the cross with attached strings) along a prescribed learning path (light gray line) consisting of knowledge bits (darker gray dots) in the fixed knowledge space (circle). Learner A, not bound to the fixed knowledge space, is self-directed and freely explores the educational field, leaving a record of past activities (black dotted line). Like the kittens, only learner A would develop agency and self-regulated learning skills.
}
    \label{fig:agency}
\end{figure}

\subsection{Engagement and motivation from gamification}
Decades of research have shown that learning occurs when the learner’s mind is actively engaged in the learning process (Dehaene, 2020). Broadly, to activate their minds, learners need to be behaviorally and cognitively engaged in their actions: taking notes while reading, writing summaries, participating in discussions, and solving problems (Bonwell \& Eison, 1991). According to the ICAP framework, the more cognitively and behaviorally engaged learners are during learning, the more they learn (Chi \& Wylie, 2014). Some adaptive learning systems, such as Korbit, promote active learning by alternating instruction with interactive problem-solving exercises, coding exercises, project-based learning, Socratic tutoring, and video lectures (St-Hilaire et al., 2022). However, a common misconception in the adaptive learning literature is that these systems must incorporate gamification, the application of game design elements such as leaderboards, badges, levels, and points, in non-game settings to increase engagement (Dicheva et al., 2015; Maghsudi et al., 2021).

In this context, an important distinction has been made between extrinsic and intrinsic motivation (Buckley \& Doyle, 2016; Mekler et al., 2017; Ryan \& Deci, 2000). Extrinsic motivation stems from extrinsic rewards (e.g., badges, points, or even grades), whereas intrinsic motivation is an inherent interest and enjoyment in anticipating and obtaining new information (Ryan \& Deci, 2000). Decades of research have shown that students must be intrinsically motivated to be engaged for extended periods (Deci et al., 2001; Oudeyer et al., 2016). Furthermore, direct attempts to control learning outcomes through extrinsic rewards and evaluations typically result in lower-quality motivation and performance (Ryan \& Deci, 2020).

In principle, when designed according to the principles of cognitive psychology, educational games can effectively support intrinsic motivation by directing children’s attention to the content and stimuli (Dehaene, 2020). Games can give students the freedom to fail, choose their next actions, provide rapid feedback, and often offer social engagement, all of which support learning (Dicheva et al., 2015). For example, adding quests to a foreign language course elicited long-term intrinsic motivation and made the course more enjoyable, challenging, and meaningful (Philpott \& Son, 2022). However, the effect of game elements on learner motivation in educational settings varies greatly depending on the specific gamification element (Sailer \& Homner, 2020).

Firstly, much evidence suggests that rewards, incentives, and competition can decrease intrinsic motivation (Deci et al., 2001; Hanus \& Fox, 2015; Ryan \& Deci, 2020). For example, a comparison of a social networking site and a gamified learning management system found that gamification increased competition and decreased student collaboration, sharing, and participation (de-Marcos et al., 2014). Courses gamified with leaderboards and badges have been shown to lower final exam scores and decrease intrinsic motivation compared to non-gamified classes (Hanus \& Fox, 2015). An experiment examining the specific effects of individual gamification elements showed that points, leaderboards, and levels did not affect intrinsic motivation and only increased the performance quantity (Mekler et al., 2017). A meta-analysis found psychological effects of gamification, such as increasing enjoyment, fostering enthusiasm, and satisfying learners’ need for recognition while providing no additional benefit and sometimes causing anxiety and jealousy (Bai et al., 2020). In summary, some gamification elements, such as leaderboards and badges, may decrease intrinsic motivation while providing no additional benefit.

Secondly, while there is sufficient evidence that gamification impacts learning outcomes, there is a paucity of data on the long-term benefits of gamification applied in educational settings (Dichev \& Dicheva, 2017). At the short-term performance level, two recent meta-analyses found a positive medium correlation between gamification and learning outcomes (Huang et al., 2020; Sailer \& Homner, 2020). However, the impact of individual gamification elements on learning has been less studied (Dicheva et al., 2015). One of the more recent meta-analyses showed how including game fiction/storytelling and collaborative social interactions has a small positive effect on behavioral, but not cognitive and motivational, learning outcomes (Sailer \& Homner, 2020). The second meta-analysis showed large effect sizes when collaboration and quests/missions/modules were included in the intervention and more significant effects when some elements, such as leaderboards and timed activities, were excluded (Huang et al., 2020).

Taken together, a scarcity of theoretical foundations marks the application of gamification in education, inconsistent empirical findings, and poor experimental design (Seaborn \& Fels, 2015). In addition, a recent longitudinal study showed that even well-designed educational games may not have sustained benefits for educational goals (Potier Watkins \& Dehaene, 2023). Hence, designers of PL technologies need to exercise caution when considering integrating both gamification elements and games into computer-based learning environments. As an alternative to gamification, solid evidence supports active learning methods' effectiveness in eliciting intrinsic motivation and enhancing deep understanding, motivation, and creativity (Chi \& Wylie, 2014).

\subsection{Individualization}

Adaptive learning systems aspire to free the learners from the shackles of space and time. In this paradigm, learners interact individually with an AI-driven learning system that sequences learning objectives and paces instruction according to individual needs (Figure 2A). Tailoring learning content to individual aptitude (i.e., provide learning content in the zone of proximal development) has been shown to increase scores and reduce study time compared to group-level instruction (Cui et al., 2018; St-Hilaire et al., 2022; Tetzlaff et al., 2021). However, while adaptive learning systems increase the efficiency of learning domain knowledge and skills, we question the power of these systems to facilitate collaborative and social learning settings. The learning-centered paradigm stems from social constructivist theory, which posits that learning occurs through social interactions (Hendrick, 2020; Palincsar, 1998; Vygotsky \& Cole, 1978). Therefore, learning requires shared knowledge construction (Hadwin et al., 2018). Although AI has been used to support collaborative learning processes (Tan et al., 2022; Ouyang et al., 2023) and strong theoretical and practical foundations have been laid by the field of Computer-Supported Collaborative Learning (Rosé \& Ferschke, 2016; Hernández-Sellés et al., 2019; Jeong et al., 2019), AI-driven PL systems are specifically designed to empower independent problem-solving and free learning from space and time constraints (Biswas et al., 2016). Face-to-face collaborative learning is seldom supported by PL systems (Han et al., 2021). Critically, the more holistic paradigm argues that PL is not about separating students to learn alone at their preferred pace (Miliband, 2006). Instead, the desired learning is collaborative and exercised in social contexts with peers, teachers, families, and communities (Figure 2B) (OECD, 2019c; Walkington \& Bernacki, 2020).

\begin{figure}
    \centering
    \includegraphics[width=1\linewidth]{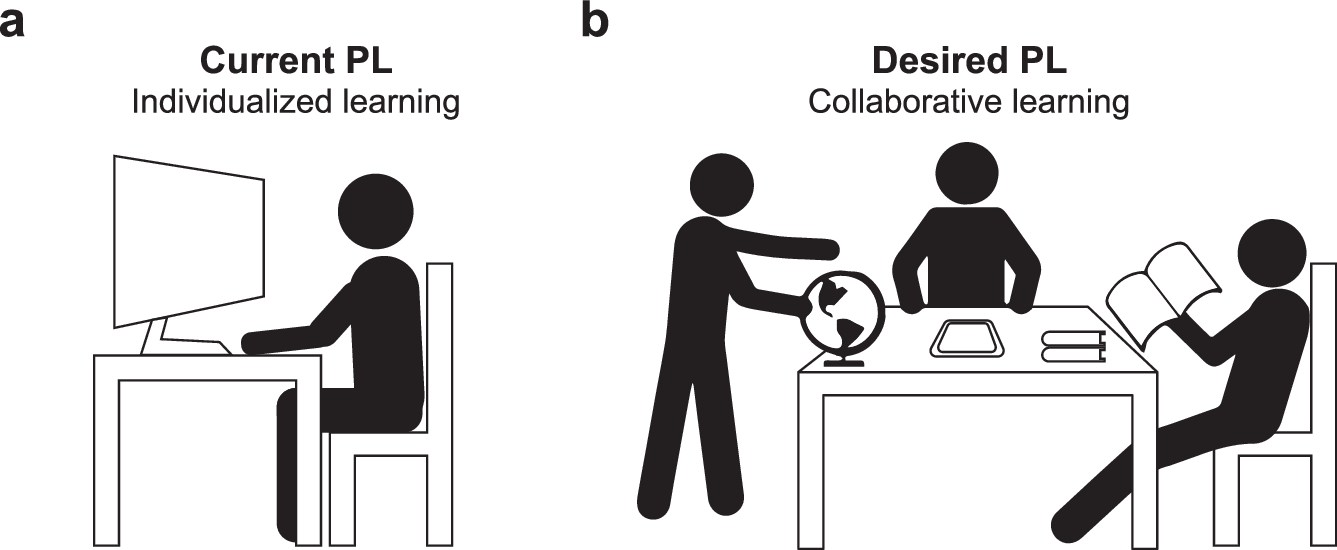}
    \caption{\textbf{a} The prevailing view of personalized learning (PL) as “individualization,” where an adaptive learning system tailors learning content to individual learners to improve their performance. This approach is mainly effective in improving the learning of subject-based knowledge. \textbf{b} The goals of modern education require a collaborative learning environment where learners can share ideas, receive and give feedback, and co-regulate. Figure adapted from (Winter, 2018) with permission from John Wiley and Sons.}
    \label{fig:individualization}
\end{figure}

\section{Beyond adaptive learning systems}
Our analysis of adaptive learning systems reveals a mismatch with the educational goals set out in the OECD Learning Compass 2030 (OECD, 2019b). While adaptive learning approaches effectively foster foundational knowledge and skills, they are limited in enhancing learners’ agency, general competencies and activating learners’ minds. However, recent advances of generative AI (GenAI) may address some of these issues and enable a new perspective on personalized learning (Kasneci et al., 2023).

\subsection{Potential of generative artificial intelligence}
The potential of GenAI could be asserted to the research on intelligent tutoring systems, a sub-category of adaptive learning systems that provide personalized instruction through natural language (speech or text) dialogues inspired by human tutoring strategies (VanLehn, 2006; Nye et al., 2014). As pointed out above, some ITS solutions are specifically designed to foster more general competencies (Benvenuti et al., 2023), SRL and metacognition (Duffy \& Azevedo, 2015; Jones \& Castellano, 2017; Azevedo et al., 2022; Roll et al., 2007; Daradoumis \& Arguedas, 2020), especially when accompanied by open learner models that tacitly invite learners to regulate learning (Long \& Aleven, 2017; Winne, 2021). Other ITS solutions have been shown to improve argumentation skills (Wambsganss et al., 2021) and facilitate collaborative learning (Haq et al., 2020; Diziol et al., 2010; Han et al., 2021).

Extending the promising potential of GenAI in education, their domain-agnostic architecture and natural language comprehension offer opportunities to address the limitations of ITS and facilitate a more general personal tutoring (Jeon \& Lee, 2023; Kasneci et al., 2023). Specifically, ITS has been using curated content and predefined dialogues (e.g., Azevedo et al., 2022), which could be improved by having more dynamic content and feedback using GenAI (e.g., Abdelghani, Wang, et al., 2022). With proper design, it may be possible that GenAI-based systems could promote learners’ SRL skills and conceptual understanding (Abdelghani, Oudeyer et al., 2022). Similarly to specific ITS solutions, GenAI solutions could act as pedagogical agents that prompt learners to engage in specific cognitive and metacognitive self-regulation strategies, expanding learning experiences beyond narrow knowledge acquisition (Dever et al., 2022; Azevedo et al., 2022; Wu et al., 2023).

Nevertheless, GenAI may not bring the benefits expected (Laak et al. 2024). Frequent GenAI usage has been shown to be negatively correlated with critical thinking abilities, with cognitive offloading acting as a mediating factor (Gerlich, 2025). Critically, ChatGPT promotes metacognitive laziness as learners become dependent on GenAI assistance, diminishing their own engagement in metacognitive processes (Fan et al., 2024). A recent meta-analysis of empirical studies on ChatGPT impact on learning showed increased performance but diminished mental effort, questioning its effects on learning (Deng et al., 2024).

The existing research on GenAI already supports the claim that to gain educational value from these models, they need to be adapted carefully and thoughtfully in the service of evidence-based learning practices (Mollick \& Mollick, 2023). For example, a GPT-3-driven conversational agent adapted to help learners generate more divergent questions increased their intrinsic motivation compared to manually predefined curiosity-prompting systems (Abdelghani, Wang et al., 2022). Another study showed how to integrate ChatGPT into realistic K-12 classroom learning settings where students are positioned as active agents in their learning, building upon each other’s ideas to advance collective understanding and solve authentic problems (Chen \& Zhu, 2023). GenAI assistants could play a learning-enhancing role if they focus on activating learners’ minds, developing learners’ self-regulation skills, and supporting collaborative learning rather than being a homework-solving crutch (Wang et al., 2024). Nevertheless, the dream of GenAI providing everyone a personal tutor has not yet materialised, primarily because of difficulties with verbalizing the complexities of effective human pedagogical intuition into GenAI prompts (Jurenka et al., 2024).

\subsection{Towards a hybrid model}
The future of formal education is likely to be a hybrid model of reciprocal interactions between humans and AI rather than a fully automated system (Molenaar, 2022b). In order to effectively self-direct their learning and keep learning throughout life, learners must have the relevant SRL skills that enable them to learn independently (Sutarni et al., 2021; Theobald, 2021; van Alten et al., 2020). The paramount importance of SRL skills was exposed during distance learning prompted by the COVID-19 pandemic. Findings indicate a positive correlation between an individual's self-regulation skills and satisfaction, well-being, performance, and self-efficacy during distance learning (Jia, 2022; Peteros et al., 2022; Sutarni et al., 2021). We recognize that a system-wide effort to support SRL is needed for the hybrid model (Hopkins, 2010; Molenaar, 2022b). Resources should be directed towards training children in SRL skills as early as possible, preferably in the preschool years, which is the most influential period for such training (Dignath et al., 2008; Kochanska et al., 2001; Moffitt et al., 2011; Muir et al., 2023). A promising way to support SRL in primary education is using written learning guides that enables self-paced learning, focuses on developing deep conceptual understanding, fosters learners’ agency through rich choice of options to control their performance etc., checking all the boxes of the Learning Compass 2030 (Kersna et al., 2025).

Because SRL skills develop with age (Panadero, 2017), the younger the learner, the more collaborative and socially rooted the learning environment should be. Conversely, the older the learner, the more effectively they could use AI assistants for their learning (Figure 3). In the early years, learning should be more social, involving interactions with peers, parents, teachers, and the community to develop SRL skills (Zimmerman, 2000). Collaborative learning is at the heart of learning-centered education and is part of developing general competencies such as critical thinking and complex problem-solving skills (Hadwin et al., 2018; Hendrick, 2020). One could imagine that because GenAI models are neither transparent nor robust, the less SRL skills a learner has the more important a teacher’s role is in detecting errors that GenAI models create (Hlosta et al., 2022). Because the need for relatedness, i.e., a sense of connection and belonging, is fundamental to human motivation, well-being, and academic performance (Ryan \& Deci, 2020), there will always be a portion of learning that needs to be grounded in social interactions (white portion of Figure 3).

\begin{figure}
    \centering
    \includegraphics[width=1\linewidth]{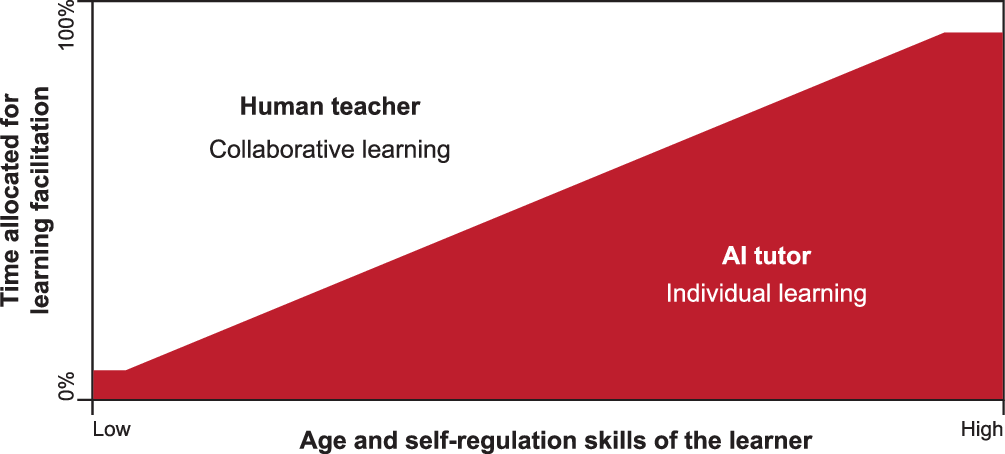}
    \caption{Human instructors are needed to support learners with developing self-regulated learning skills. Scaffolded assistance is required as a function of learner’s age: younger students require more guidance and a social learning environment, while older students require less support. AI assistants are more effective for learners who already exhibit self-regulation skills.}
    \label{fig:hybrid}
\end{figure}

\section{Discussion}

The technological approach to personalized learning has come with the promise of equitable access to high-quality education. This paper analyzed the alignment of AI-driven personalized learning systems with the goals of modern education to challenge this promise. We specifically asked how this technological approach to PL adheres to the goals set out in the OECD Learning Compass 2030: fostering students’ agency and self-regulation skills, developing general competencies, and cognitive activation of the learners’ minds (OECD, 2019b).

We explored this question by highlighting five challenges the technological approach to PL has in relation to these educational goals. First, the goal of education is not to improve student performance on post-test scores but to foster long-term conceptual understanding through methods that cognitively engage with the learner’s mind. Second, adaptive learning systems are limited to predefined content models, hindering learners’ creativity to think outside of the predefined knowledge set. Third, students should have the opportunity to construct their own learning paths and exert agency over their learning choices in order to develop healthy self-regulated learning skills required for lifelong learning. In addition, AI-driven PL systems often rely on gamification as a source of engagement and motivation instead of activating the learners’ mind through well-known active learning methods - Fourth. Finally, adaptive learning systems lead to highly individualized learning experiences, whereas the aspired learning environment should be collaborative and social. This misalignment is summarized in Table 1 below.

\begin{table}[H]
    \caption{Alignment of modern education and AI-driven personalized learning systems}
    \centering
    \begin{tabular}{p{4cm} p{5cm} p{5cm}}
        \toprule
        & \textbf{Adaptive learning systems} & \textbf{Educational goal} \\
        \midrule
        \textbf{Success metric} & Increased test scores (performance) & Conceptual understanding \\
        \textbf{Learning content} & Mostly domain knowledge & Includes general competencies \\
        \textbf{Learner’s agency} & Hinders & Empowers \\
        \textbf{Type of engagement} & External from gamification & Internal from cognitive activation \\
        \textbf{Approach} & Individualistic & Collaborative \\
        \bottomrule
    \end{tabular}
    \label{tab:alignment}
\end{table}

Taken together, we highlight a misalignment between the educational goals outlined in the OECD Learning Compass 2030 and the AI-driven approach to personalized learning. Although adaptive learning systems have proven effective for acquiring domain knowledge, their limitations could hinder the development of learners’ agency, general competencies, and SRL skills. The narrow view of considering education as an individual information gathering and declarative knowledge acquisition process is elegantly challenged by Cukurova (2024): “Learning is not only about absorbing information and education is not only about learning”. While the impetus behind the technology-based approach to PL is motivated by Bloom's 2-sigma problem (1984), the aspirations of modern education have moved beyond, requiring the development of broader competencies and skills necessary for lifelong learning, transfer, and well-being. From that perspective, the technological approach to PL is yet to satisfy the modern standards of high-quality education, and could only be viewed as a more efficient version of the “traditional” education. It has been argued that the current education system has not been a one-size-fits-all approach for a long time, and anyone arguing against it is forming an artificial cause for a reason that requires further analysis (Pelletier, 2024). Without completely abandoning the idea of a more personalized approach to education, we ought to ask what is personal about learning (Laak et al., 2024) and what kind of people we need for our societies to flourish (Tuomi, 2023).

There could be many reasons why the current approach to PL is limited to a technological approach. For example, it may be that it is simply more profitable for the multi-billion dollar market of AIEd (Guan et al., 2020; Holmes et al., 2019; Miao et al., 2021) to apply existing recommendation systems (e.g., those in YouTube, Amazon, TikTok, etc.) rather than investing in understanding and solving serious educational needs and problems (Holmes et al., 2022; Reeves \& Lin, 2020; Zawacki-Richter et al., 2019). We are concerned that applying AI without a deeper consideration of its educational value could slow educational progress by diverting focus and resources from scientifically proven changes that need to be made in education systems. For example, as we have argued above, moving learners along a prescribed learning path can rob them of opportunities for playful discovery, limit their agency, and ultimately become a barrier to innovation, creativity, and even happiness (Stanley \& Lehman, 2015). As noted more than a decade ago, AIEd should start with education goals and then choose the appropriate technology to support PL, not the other way around (Vandewaetere \& Clarebout, 2014).

\section{Conclusion}
The technological approach to personalized learning offers notable advantages in terms of efficiency and individual pacing in domain-specific knowledge acquisition, yet it struggles to align with the broader goals of modern education. As highlighted in our analysis, this approach prioritizes individualization and performance-based assessment over more holistic educational goals, such as fostering learner agency, cognitive engagement and self-regulated learning skills, and developing general competencies. While carefully designed generative AI solutions could enhance some of these goals, emerging research highlights the potential negative effects of unregulated use of generative AI on learning, mental effort, and metacognitive abilities. The future is possibly a human-AI hybrid model, where thoughtfully implemented AI serves as a supportive role within collaborative, socially grounded learning environments, presenting a more promising avenue for aligning technology with the needs of contemporary education.

\section*{References}
Abdelghani, R., Oudeyer, P.-Y., Law, E., de Vulpillières, C., \& Sauzéon, H. (2022). Conversational agents for fostering curiosity-driven learning in children. \textit{International Journal of Human-Computer Studies}, \textit{167}, 102887. https://doi.org/10.1016/j.ijhcs.2022.102887 

Abdelghani, R., Wang, Y.-H., Yuan, X., Wang, T., Sauzéon, H., \& Oudeyer, P.-Y. (2022). \textit{GPT-3-driven pedagogical agents for training children’s curious question-asking skills} (arXiv:2211.14228). arXiv. https://doi.org/10.48550/arXiv.2211.14228 

Aleven, V., McLaren, B., Roll, I., \& Koedinger, K. (2006). Toward Meta-cognitive Tutoring: A Model of Help Seeking with a Cognitive Tutor. \textit{International Journal of Artificial Intelligence in Education}, \textit{16}(2), 101–128.

Azevedo, R., \& Aleven, V. (Eds.). (2013). \textit{International Handbook of Metacognition and Learning Technologies} (Vol. 28). Springer. https://doi.org/10.1007/978-1-4419-5546-3

Azevedo, R., Bouchet, F., Duffy, M., Harley, J., Taub, M., Trevors, G., Cloude, E., Dever, D., Wiedbusch, M., Wortha, F., \& Cerezo, R. (2022). Lessons Learned and Future Directions of MetaTutor: Leveraging Multichannel Data to Scaffold Self-Regulated Learning With an Intelligent Tutoring System. \textit{Frontiers in Psychology}, \textit{13}.

Azevedo, R., Witherspoon, A., Chauncey, A., Burkett, C., \& Fike, A. (2009). MetaTutor: A MetaCognitive tool for enhancing self-regulated learning. \textit{AAAI Fall Symposium Series}.

Bai, S., Hew, K. F., \& Huang, B. (2020). Does gamification improve student learning outcome? Evidence from a meta-analysis and synthesis of qualitative data in educational contexts. \textit{Educational Research Review}, \textit{30}, 100322. https://doi.org/10.1016/j.edurev.2020.100322 

Bail, F. T., Zhang, S., \& Tachiyama, G. T. (2008). Effects of a Self-Regulated Learning Course on the Academic Performance and Graduation Rate of College Students in an Academic Support Program. \textit{Journal of College Reading and Learning}, \textit{39}(1), 54–73. https://doi.org/10.1080/10790195.2008.10850312

Barnard-Brak, L., Lan, W. Y., \& Paton, V. O. (2010). Profiles in Self-Regulated Learning in the Online Learning Environment. \textit{International Review of Research in Open and Distributed Learning}, \textit{11}(1), 1–181.

Bartl, E., \& Belohlavek, R. (2011). Knowledge spaces with graded knowledge states. \textit{Information Sciences}, \textit{181}(8), 1426–1439. https://doi.org/10.1016/j.ins.2010.11.040 

Beese, E. B. (2019). A process perspective on research and design issues in educational personalization. \textit{Theory and Research in Education}, \textit{17}(3), 253–279. https://doi.org/10.1177/1477878519893963 

Benvenuti, M., Cangelosi, A., Weinberger, A., Mazzoni, E., Benassi, M., Barbaresi, M., \& Orsoni, M. (2023). Artificial intelligence and human behavioral development: A perspective on new skills and competences acquisition for the educational context. \textit{Computers in Human Behavior}, \textit{148}, 107903. \href{https://doi.org/10.1016/j.chb.2023.107903}{https://doi.org/10.1016/j.chb.2023.107903}

Bernacki, M. L., Greene, M. J., \& Lobczowski, N. G. (2021). A Systematic Review of Research on Personalized Learning: Personalized by Whom, to What, How, and for What Purpose(s)? \textit{Educational Psychology Review}, \textit{33}(4), 1675–1715. https://doi.org/10.1007/s10648-021-09615-8

Bertschinger, N., Olbrich, E., Ay, N., \& Jost, J. (2008). Autonomy: An information theoretic perspective. \textit{Biosystems}, \textit{91}(2), 331–345. https://doi.org/10.1016/j.biosystems.2007.05.018

Biswas, G., Leelawong, K., Schwartz, D., Vye, N., \& The Teachable Agents Group at Vanderbilt. (2005). Learning by Teaching: A New Agent Paradigm for Educational Software. \textit{Applied Artificial Intelligence}, \textit{19}(3–4), 363–392. https://doi.org/10.1080/08839510590910200

Biswas, G., Segedy, J. R., \& Bunchongchit, K. (2016). From Design to Implementation to Practice a Learning by Teaching System: Betty’s Brain. \textit{International Journal of Artificial Intelligence in Education}, \textit{26}(1), 350–364. https://doi.org/10.1007/s40593-015-0057-9

Bloom, B. S. (1984). The 2 Sigma Problem: The Search for Methods of Group Instruction as Effective as One-to-One Tutoring. \textit{Educational Researcher}, \textit{13}(6), 4–16. https://doi.org/10.3102/0013189X013006004 

Bonwell, C. C., \& Eison, J. A. (1991). \textit{Active Learning: Creating Excitement in the Classroom. 1991 ASHE-ERIC Higher Education Reports}. ERIC Clearinghouse on Higher Education, The George Washington University, One Dupont Circle, Suite 630, Washington, DC 20036-1183 (\$17. https://eric.ed.gov/?id=ED336049 

Bransford, J., National Research Council (U.S.), \& National Research Council (U.S.) (Eds.). (2000). \textit{How people learn: Brain, mind, experience, and school} (Expanded ed). National Academy Press. 

Brod, G., Hasselhorn, M., \& Bunge, S. A. (2018). When generating a prediction boosts learning: The element of surprise. \textit{Learning and Instruction}, \textit{55}, 22–31. https://doi.org/10.1016/j.learninstruc.2018.01.013 

Brod, G., Kucirkova, N., Shepherd, J., Jolles, D., \& Molenaar, I. (2023). Agency in Educational Technology: Interdisciplinary Perspectives and Implications for Learning Design. \textit{Educational Psychology Review}, \textit{35}(1), 25. https://doi.org/10.1007/s10648-023-09749-x 

Buckley, P., \& Doyle, E. (2016). Gamification and student motivation. \textit{Interactive Learning Environments}, \textit{24}(6), 1162–1175. https://doi.org/10.1080/10494820.2014.964263 

Capuano, N., \& Caballé, S. (2020). Adaptive Learning Technologies. \textit{AI Magazine}, \textit{41}(2), Article 2. https://doi.org/10.1609/aimag.v41i2.5317 

Chen, X., Xie, H., Zou, D., \& Hwang, G.-J. (2020). Application and theory gaps during the rise of Artificial Intelligence in Education. \textit{Computers and Education: Artificial Intelligence}, \textit{1}, 100002. \href{https://doi.org/10.1016/j.caeai.2020.100002}{https://doi.org/10.1016/j.caeai.2020.100002}

Chen, B., \& Zhu, X. (2023). Integrating generative AI in knowledge building. \textit{Computers and Education: Artificial Intelligence}, 100184. https://doi.org/10.1016/j.caeai.2023.100184

Chi, M. T. H., \& Wylie, R. (2014). The ICAP Framework: Linking Cognitive Engagement to Active Learning Outcomes. \textit{Educational Psychologist}, \textit{49}(4), 219–243. https://doi.org/10.1080/00461520.2014.965823 

Corbett, A. T., Koedinger, K. R., \& Anderson, J. R. (1997). Intelligent Tutoring Systems. In M. G. Helander, T. K. Landauer, \& P. V. Prabhu (Eds.), \textit{Handbook of Human-Computer Interaction (Second Edition)} (pp. 849–874). North-Holland. https://doi.org/10.1016/B978-044481862-1.50103-5 

Council of the EU. (2002). \textit{Council Resolution of 27 June 2002 on lifelong learning}. Publications Office of the EU. https://op.europa.eu/en/publication-detail/-/publication/0bf0f197-5b35-4a97-9612-19674583cb5b 

Cui, W., Xue, Z., \& Thai, K.-P. (2018). Performance Comparison of an AI-Based Adaptive Learning System in China. \textit{2018 Chinese Automation Congress (CAC)}, 3170–3175. https://doi.org/10.1109/CAC.2018.8623327 

Cukurova, M., Miao, X., \& Brooker, R. (2023). Adoption of Artificial Intelligence in Schools: Unveiling Factors Influencing Teachers’ Engagement. In N. Wang, G. Rebolledo-Mendez, N. Matsuda, O. C. Santos, \& V. Dimitrova (Eds.), \textit{Artificial Intelligence in Education} (pp. 151–163). Springer Nature Switzerland. https://doi.org/10.1007/978-3-031-36272-9\_13

Daradoumis, T., \& Arguedas, M. (2020). Cultivating Students’ Reflective Learning in Metacognitive Activities through an Affective Pedagogical Agent. \textit{Journal of Educational Technology \& Society}, \textit{23}(2), 19–31. 

Dever, D. A., Sonnenfeld, N. A., Wiedbusch, M. D., \& Azevedo, R. (2022). Pedagogical Agent Support and Its Relationship to Learners’ Self-regulated Learning Strategy Use with an Intelligent Tutoring System. In M. M. Rodrigo, N. Matsuda, A. I. Cristea, \& V. Dimitrova (Eds.), \textit{Artificial Intelligence  in Education} (pp. 332–343). Springer International Publishing. https://doi.org/10.1007/978-3-031-11644-5\_27

de-Marcos, L., Domínguez, A., Saenz-de-Navarrete, J., \& Pagés, C. (2014). An empirical study comparing gamification and social networking on e-learning. \textit{Computers \& Education}, \textit{75}, 82–91. https://doi.org/10.1016/j.compedu.2014.01.012 

Deci, E. L., Koestner, R., \& Ryan, R. M. (2001). Extrinsic Rewards and Intrinsic Motivation in Education: Reconsidered Once Again. \textit{Review of Educational Research}, \textit{71}(1), 1–27. https://doi.org/10.3102/00346543071001001 

Dehaene, S. (2020). \textit{How We Learn: The New Science of Education and the Brain}. Penguin UK. 

Dempster, F. N., \& Farris, R. (1990). The spacing effect: Research and practice. \textit{Journal of Research \& Development in Education}, \textit{23}, 97–101. 

Deng, Ruiqi, Maoli Jiang, Xinlu Yu, Yuyan Lu, and Shasha Liu. Does ChatGPT enhance student learning? A systematic review and meta-analysis of experimental studies. \textit{Computers \& Education} (2024): 105224. https://doi.org/10.1016/j.compedu.2024.105224

Dichev, C., \& Dicheva, D. (2017). Gamifying education: What is known, what is believed and what remains uncertain: a critical review. \textit{International Journal of Educational Technology in Higher Education}, \textit{14}(1), 9. https://doi.org/10.1186/s41239-017-0042-5 

Dicheva, D., Dichev, C., Agre, G., \& Angelova, G. (2015). Gamification in Education: A Systematic Mapping Study. \textit{Journal of Educational Technology \& Society}, \textit{18}(3), 75–88. 

Dignath, C., Buettner, G., \& Langfeldt, H.-P. (2008). How can primary school students learn self-regulated learning strategies most effectively?: A meta-analysis on self-regulation training programmes. \textit{Educational Research Review}, \textit{3}(2), 101–129. https://doi.org/10.1016/j.edurev.2008.02.003 

Dignath, C., \& Veenman, M. V. J. (2021). The Role of Direct Strategy Instruction and Indirect Activation of Self-Regulated Learning—Evidence from Classroom Observation Studies. \textit{Educational Psychology Review}, \textit{33}(2), 489–533.\href{https://doi.org/10.1007/s10648-020-09534-0}{https://doi.org/10.1007/s10648-020-09534-0}

Diziol, D., Walker, E., Rummel, N., \& Koedinger, K. R. (2010). Using Intelligent Tutor Technology to Implement Adaptive Support for Student Collaboration. \textit{Educational Psychology Review}, \textit{22}(1), 89–102. https://doi.org/10.1007/s10648-009-9116-9

Doignon, J.-P., \& Falmagne, J.-C. (2012). \textit{Knowledge Spaces}. Springer Science \& Business Media. 

du Boulay, B. (2019). Escape from the Skinner Box: The case for contemporary intelligent learning environments. \textit{British Journal of Educational Technology}, \textit{50}(6), 2902–2919. https://doi.org/10.1111/bjet.12860

Duffy, M. C., \& Azevedo, R. (2015). Motivation matters: Interactions between achievement goals and agent scaffolding for self-regulated learning within an intelligent tutoring system. \textit{Computers in Human Behavior}, \textit{52}, 338–348. https://doi.org/10.1016/j.chb.2015.05.041

Essa, A. (2016). A possible future for next generation adaptive learning systems. \textit{Smart Learning Environments}, \textit{3}(1), 16. https://doi.org/10.1186/s40561-016-0038-y 

Ezzaim, A., Kharroubi, F., Dahbi, A., Aqqal, A., \& Haidine, A. (2022). Artificial intelligence in education—State of the art. \textit{International Journal of Computer Engineering and Data Science (IJCEDS)}, \textit{2}(2), Article 2. http://www.ijceds.com/ijceds/article/view/37 

Fan, Y., Tang, L., Le, H., Shen, K., Tan, S., Zhao, Y., ... \& Gašević, D. (2024). Beware of metacognitive laziness: Effects of generative artificial intelligence on learning motivation, processes, and performance. \textit{British Journal of Educational Technology}. \href{https://doi.org/10.1111/bjet.13544}{https://doi.org/10.1111/bjet.13544}

Feng, M., Cui, W., \& Wang, S. (2018). Adaptive Learning Goes to China. In C. Penstein Rosé, R. Martínez-Maldonado, H. U. Hoppe, R. Luckin, M. Mavrikis, K. Porayska-Pomsta, B. McLaren, \& B. du Boulay (Eds.), \textit{Artificial Intelligence in Education} (pp. 89–93). Springer International Publishing. https://doi.org/10.1007/978-3-319-93846-2\_17 

Gerbier, E., \& Toppino, T. C. (2015). The effect of distributed practice: Neuroscience, cognition, and education. \textit{Trends in Neuroscience and Education}, \textit{4}(3), 49–59. https://doi.org/10.1016/j.tine.2015.01.001 

Gerlich, M. (2025). AI Tools in Society: Impacts on Cognitive Offloading and the Future of Critical Thinking. \textit{Societies}, 15(1), 6. https://doi.org/10.3390/soc15010006

Greer, J. E., \& McCalla, G. I. (1991). \textit{Student Modelling: The Key to Individualized Knowledge-Based Instruction}. Springer Science \& Business Media. 

Groff, J. S. (2017). The State of the Field \& Future Directions. \textit{Center for Curriculum Redesign}, 47. 

Guan, C., Mou, J., \& Jiang, Z. (2020). Artificial intelligence innovation in education: A twenty-year data-driven historical analysis. \textit{International Journal of Innovation Studies}, \textit{4}(4), 134–147. https://doi.org/10.1016/j.ijis.2020.09.001 

Hacker, D. J., Dunlosky, J., \& Graesser, A. C. (2009). \textit{Handbook of Metacognition in Education}. Routledge.

Hadwin, A., Järvelä, S., \& Miller, M. (2018). Self-regulation, co-regulation, and shared regulation in collaborative learning environments. In \textit{Handbook of self-regulation of learning and performance, 2nd ed} (pp. 83–106). Routledge/Taylor \& Francis Group. https://doi.org/10.4324/9781315697048-6 

Han, J., Kim, K. H., Rhee, W., \& Cho, Y. H. (2021). Learning analytics dashboards for adaptive support in face-to-face collaborative argumentation. \textit{Computers \& Education}, \textit{163}, 104041. https://doi.org/10.1016/j.compedu.2020.104041

Hanus, M. D., \& Fox, J. (2015). Assessing the effects of gamification in the classroom: A longitudinal study on intrinsic motivation, social comparison, satisfaction, effort, and academic performance. \textit{Computers \& Education}, \textit{80}, 152–161. https://doi.org/10.1016/j.compedu.2014.08.019 

Hattie, J. (2011). \textit{Visible Learning for Teachers: Maximizing Impact on Learning}. Routledge. https://doi.org/10.4324/9780203181522 

Hattie, J., \& Timperley, H. (2007). The Power of Feedback. \textit{Review of Educational Research}, \textit{77}(1), 81–112. https://doi.org/10.3102/003465430298487 

Haq, I. U., Anwar, A., Basharat, I., \& Sultan, K. (2020). Intelligent Tutoring Supported Collaborative Learning (ITSCL): A Hybrid Framework. \textit{International Journal of Advanced Computer Science and Applications (IJACSA)}, \textit{11}(8), Article 8. https://doi.org/10.14569/IJACSA.2020.0110866

Harati, H., Sujo-Montes, L., Tu, C.-H., Armfield, S. J. W., \& Yen, C.-J. (2021). Assessment and Learning in Knowledge Spaces (ALEKS) Adaptive System Impact on Students’ Perception and Self-Regulated Learning Skills. \textit{Education Sciences}, \textit{11}(10), Article 10. https://doi.org/10.3390/educsci11100603

Held, R., \& Hein, A. (1963). Movement-produced stimulation in the development of visually guided behavior. \textit{Journal of Comparative and Physiological Psychology}, \textit{56}, 872–876. https://doi.org/10.1037/h0040546 

Hendrick, P. K., Carl. (2020). \textit{How Learning Happens: Seminal Works in Educational Psychology and What They Mean in Practice}. Routledge. https://doi.org/10.4324/9780429061523 

Hernández-Sellés, N., Pablo-César Muñoz-Carril, \& González-Sanmamed, M. (2019). Computer-supported collaborative learning: An analysis of the relationship between interaction, emotional support and online collaborative tools. \textit{Computers \& Education}, \textit{138}, 1–12. https://doi.org/10.1016/j.compedu.2019.04.012

Hinojo-Lucena, F.-J., Aznar-Díaz, I., Cáceres-Reche, M.-P., \& Romero-Rodríguez, J.-M. (2019). Artificial Intelligence in Higher Education: A Bibliometric Study on its Impact in the Scientific Literature. \textit{Education Sciences}, \textit{9}(1), Article 1. https://doi.org/10.3390/educsci9010051 

Hintzman, D. L. (1974). Theoretical implications of the spacing effect. In \textit{Theories in cognitive psychology: The Loyola Symposium} (pp. xi, 386–xi, 386). Lawrence Erlbaum. 

Hlosta, M., Herodotou, C., Papathoma, T., Gillespie, A., \& Bergamin, P. (2022). Predictive learning analytics in online education: A deeper understanding through explaining algorithmic errors. \textit{Computers and Education: Artificial Intelligence}, 100108. https://doi.org/10.1016/j.caeai.2022.100108

Hofer, B. K., \& Yu, S. L. (2003). Teaching Self-Regulated Learning Through a ‘Learning to Learn’ Course. \textit{Teaching of Psychology}, \textit{30}(1), 30–33. https://doi.org/10.1207/S15328023TOP3001\_05

Holmes, W., Bialik, M., \& Fadel, C. (2019). \textit{Artificial intelligence in education: Promises and implications for teaching and learning}. 

Holmes, W., Persson, J., Chounta, I.-A., Wasson, B., \& Dimitrova, V. (2022). \textit{Artificial intelligence and education: A critical view through the lens of human rights, democracy and the rule of law}. Council of Europe. 

Hopkins, D. (2010). Personalized Learning in School Age Education. In \textit{International Encyclopedia of Education} (pp. 227–232). Elsevier. https://doi.org/10.1016/B978-0-08-044894-7.01073-3 

Huang, R., Ritzhaupt, A. D., Sommer, M., Zhu, J., Stephen, A., Valle, N., Hampton, J., \& Li, J. (2020). The impact of gamification in educational settings on student learning outcomes: A meta-analysis. \textit{Educational Technology Research and Development}, \textit{68}(4), 1875–1901. https://doi.org/10.1007/s11423-020-09807-z 

Järvelä, S., Malmberg, J., Haataja, E., Sobocinski, M., \& Kirschner, P. A. (2021). What multimodal data can tell us about the students’ regulation of their learning process? \textit{Learning and Instruction}, \textit{72}, 101203. https://doi.org/10.1016/j.learninstruc.2019.04.004 

Jarvis, P. (2004). \textit{Adult Education and Lifelong Learning: Theory and Practice} (3rd ed.). Routledge. https://doi.org/10.4324/9780203561560 

Jeon, J., \& Lee, S. (2023). Large language models in education: A focus on the complementary relationship between human teachers and ChatGPT. \textit{Education and Information Technologies}. https://doi.org/10.1007/s10639-023-11834-1 

Jeong, H., Hmelo-Silver, C. E., \& Jo, K. (2019). Ten years of Computer-Supported Collaborative Learning: A meta-analysis of CSCL in STEM education during 2005–2014. \textit{Educational Research Review}, \textit{28}, 100284. https://doi.org/10.1016/j.edurev.2019.100284

Jia, M. (2022). The influence of distance learning during COVID-19 pandemic on student’s self-regulated learning in higher education: A qualitative study. \textit{Proceedings of the 5th International Conference on Digital Technology in Education}, 47–52. https://doi.org/10.1145/3488466.3488492 

Jones, A., \& Castellano, G. (2018). Adaptive Robotic Tutors that Support Self-Regulated Learning: A Longer-Term Investigation with Primary School Children. \textit{International Journal of Social Robotics}, \textit{10}(3), 357–370. https://doi.org/10.1007/s12369-017-0458-z

Kasneci, E., Seßler, K., Küchemann, S., Bannert, M., Dementieva, D., Fischer, F., Gasser, U., Groh, G., Günnemann, S., Hüllermeier, E., Krusche, S., Kutyniok, G., Michaeli, T., Nerdel, C., Pfeffer, J., Poquet, O., Sailer, M., Schmidt, A., Seidel, T., … Kasneci, G. (2023). \textit{ChatGPT for Good? On Opportunities and Challenges of Large Language Models for Education} [Preprint]. EdArXiv. https://doi.org/10.35542/osf.io/5er8f 

Katz, S., Albacete, P., Chounta, I.-A., Jordan, P., McLaren, B. M., \& Zapata-Rivera, D. (2021). Linking Dialogue with Student Modelling to Create an Adaptive Tutoring System for Conceptual Physics. \textit{International Journal of Artificial Intelligence in Education}, \textit{31}(3), 397–445. https://doi.org/10.1007/s40593-020-00226-y 

Kerr, P. (2016). Adaptive learning. \textit{ELT Journal}, \textit{70}(1), 88–93. https://doi.org/10.1093/elt/ccv055 

Kersna, L., Laak, K. J., Lepp, L., \& Pedaste, M. (2025). Supporting Self-Regulated Learning in Primary Education: Using Written Learning Guides in the Lessons. \textit{Education Sciences}, 15(1), 60. \href{https://doi.org/10.3390/educsci15010060}{https://doi.org/10.3390/educsci15010060}

Knowles, M. S. (1975). \textit{Self-Directed Learning: A Guide for Learners and Teachers}. Association Press, 291 Broadway, New York.

Kochanska, G., Coy, K. C., \& Murray, K. T. (2001). The Development of Self-Regulation in the First Four Years of Life. \textit{Child Development}, \textit{72}(4), 1091–1111. https://doi.org/10.1111/1467-8624.00336

Koedinger, K. R., Kim, J., Jia, J. Z., McLaughlin, E. A., \& Bier, N. L. (2015). Learning is Not a Spectator Sport: Doing is Better than Watching for Learning from a MOOC. \textit{Proceedings of the Second (2015) ACM Conference on Learning @ Scale}, 111–120. https://doi.org/10.1145/2724660.2724681

Laak, K. J., Abdelghani, R., \& Aru, J. (2024, July). Personalisation is not Guaranteed: The Challenges of Using Generative AI for Personalised Learning. \textit{In International conference on innovative technologies and learning} (pp. 40-49). Cham: Springer Nature Switzerland. https://doi.org/10.1007/978-3-031-65881-5\_5

Lee, D. (2014). How to Personalize Learning in K-12 Schools: Five Essential Design Features. \textit{Educational Technology}, \textit{54}(3), 12–17. 

Lee, D., Huh, Y., Lin, C.-Y., \& Reigeluth, C. M. (2018). Technology functions for personalized learning in learner-centered schools. \textit{Educational Technology Research and Development}, \textit{66}(5), 1269–1302. https://doi.org/10.1007/s11423-018-9615-9 

Leelawong, K., \& Biswas, G. (2008). Designing Learning by Teaching Agents: The Betty’s Brain System. \textit{International Journal of Artificial Intelligence in Education}, \textit{18}(3), 181–208.

Li, H., Cui, W., Xu, Z., Zhu, Z., \& Feng, M. (2018). Yixue Adaptive Learning System and Its Promise on Improving Student Learning: \textit{Proceedings of the 10th International Conference on Computer Supported Education}, 45–52. https://doi.org/10.5220/0006689800450052 

Li, K. C., \& Wong, B. T.-M. (2021). Features and trends of personalised learning: A review of journal publications from 2001 to 2018. \textit{Interactive Learning Environments}, \textit{29}(2), 182–195. https://doi.org/10.1080/10494820.2020.1811735 

Long, Y., \& Aleven, V. (2017). Enhancing learning outcomes through self-regulated learning support with an Open Learner Model. \textit{User Modeling and User-Adapted Interaction}, \textit{27}(1), 55–88. https://doi.org/10.1007/s11257-016-9186-6

Luckin, R., \& Cukurova, M. (2019). Designing educational technologies in the age of AI: A learning sciences-driven approach. \textit{British Journal of Educational Technology}, \textit{50}(6), 2824–2838. https://doi.org/10.1111/bjet.12861 

Luckin, R., Holmes, W., Griffiths, M., Corcier, L. B., \& Pearson (Firm), U. C., London. (2016). \textit{Intelligence unleashed: An argument for AI in education}. https://www.pearson.com/content/dam/corporate/global/pearson-dot-com/files/innovation/Intelligence-Unleashed-Publication.pdf 

Ma, W., Adesope, O. O., Nesbit, J. C., \& Liu, Q. (2014). Intelligent tutoring systems and learning outcomes: A meta-analysis. \textit{Journal of Educational Psychology}, \textit{106}(4), 901–918. https://doi.org/10.1037/a0037123

Maghsudi, S., Lan, A., Xu, J., \& van der Schaar, M. (2021). Personalized Education in the Artificial Intelligence Era: What to Expect Next. \textit{IEEE Signal Processing Magazine}, \textit{38}(3), 37–50. https://doi.org/10.1109/MSP.2021.3055032 

Martin, F., Chen, Y., Moore, R. L., \& Westine, C. D. (2020). Systematic review of adaptive learning research designs, context, strategies, and technologies from 2009 to 2018. \textit{Educational Technology Research and Development}, \textit{68}(4), 1903–1929. https://doi.org/10.1007/s11423-020-09793-2 

McCabe, J. (2011). Metacognitive awareness of learning strategies in undergraduates. \textit{Memory \& Cognition}, \textit{39}(3), 462–476. https://doi.org/10.3758/s13421-010-0035-2 

Mekler, E. D., Brühlmann, F., Tuch, A. N., \& Opwis, K. (2017). Towards understanding the effects of individual gamification elements on intrinsic motivation and performance. \textit{Computers in Human Behavior}, \textit{71}, 525–534. https://doi.org/10.1016/j.chb.2015.08.048 

Miao, F., Holmes, W., Ronghuai, H., \& Hui, Z. (2021). \textit{AI and education: Guidance for policy-makers}. UNESCO. https://unesdoc.unesco.org/ark:/48223/pf0000376709 

Miliband, D. (2006). Choice and voice in personalised learning. In \textit{Schooling for tomorrow: Personalising education} (pp. 21–30). OECD Publishing.

Moffitt, T. E., Arseneault, L., Belsky, D., Dickson, N., Hancox, R. J., Harrington, H., Houts, R., Poulton, R., Roberts, B. W., Ross, S., Sears, M. R., Thomson, W. M., \& Caspi, A. (2011). A gradient of childhood self-control predicts health, wealth, and public safety. \textit{Proceedings of the National Academy of Sciences}, \textit{108}(7), 2693–2698. https://doi.org/10.1073/pnas.1010076108 

Molenaar, I. (2022a). The concept of hybrid human-AI regulation: Exemplifying how to support young learners’ self-regulated learning. \textit{Computers and Education: Artificial Intelligence}, \textit{3}, 100070. https://doi.org/10.1016/j.caeai.2022.100070 

Molenaar, I. (2022b). Towards hybrid human-AI learning technologies. \textit{European Journal of Education}, \textit{57}(4), 632–645. https://doi.org/10.1111/ejed.12527 

Mollick, E. R., \& Mollick, L. (2023). \textit{Using AI to Implement Effective Teaching Strategies in Classrooms: Five Strategies, Including Prompts} (SSRN Scholarly Paper No. 4391243). https://doi.org/10.2139/ssrn.4391243 

Muir, R. A., Howard, S. J., \& Kervin, L. (2023). Interventions and Approaches Targeting Early Self-Regulation or Executive Functioning in Preschools: A Systematic Review. \textit{Educational Psychology Review}, \textit{35}(1), 27. https://doi.org/10.1007/s10648-023-09740-6 

Munshi, A., Biswas, G., Baker, R., Ocumpaugh, J., Hutt, S., \& Paquette, L. (2023). Analysing adaptive scaffolds that help students develop self-regulated learning behaviours. \textit{Journal of Computer Assisted Learning}, \textit{39}(2), 351–368. https://doi.org/10.1111/jcal.12761

Nguyen, A., Ngo, H. N., Hong, Y., Dang, B., \& Nguyen, B.-P. T. (2022). Ethical principles for artificial intelligence in education. \textit{Education and Information Technologies}. https://doi.org/10.1007/s10639-022-11316-w 

Nye, B. D., Graesser, A. C., \& Hu, X. (2014). AutoTutor and Family: A Review of 17 Years of Natural Language Tutoring. \textit{International Journal of Artificial Intelligence in Education}, 24(4), 427–469. https://doi.org/10.1007/s40593-014-0029-5

O’Brien, J. G., Millis, B. J., \& Cohen, M. W. (2009). \textit{The Course Syllabus: A Learning-Centered Approach}. John Wiley \& Sons. 

OECD. (2006). \textit{Personalising Education}. OECD. https://doi.org/10.1787/9789264036604-en 

OECD. (2019a). \textit{Anticipation-Action-Reflection cycle for 2030}. OECD Publishing. 

OECD. (2019b). \textit{OECD Learning Compass 2030}. OECD Publishing. 

OECD. (2019c). \textit{Student agency for 2030}. OECD Publishing. 

OECD. (2019d). \textit{Transformative competencies for 2030}. OECD Publishing. 

Osakwe, I., Chen, G., Whitelock-Wainwright, A., Gašević, D., Pinheiro Cavalcanti, A., \& Ferreira Mello, R. (2022). Towards automated content analysis of educational feedback: A multi-language study. \textit{Computers and Education: Artificial Intelligence}, \textit{3}, 100059. https://doi.org/10.1016/j.caeai.2022.100059 

Oudeyer, P.-Y., Gottlieb, J., \& Lopes, M. (2016). Chapter 11 - Intrinsic motivation, curiosity, and learning: Theory and applications in educational technologies. In B. Studer \& S. Knecht (Eds.), \textit{Progress in Brain Research} (Vol. 229, pp. 257–284). Elsevier. https://doi.org/10.1016/bs.pbr.2016.05.005 

Ouyang, F., Xu, W., \& Cukurova, M. (2023). An artificial intelligence-driven learning analytics method to examine the collaborative problem-solving process from the complex adaptive systems perspective. \textit{International Journal of Computer-Supported Collaborative Learning}, \textit{18}(1), 39–66. https://doi.org/10.1007/s11412-023-09387-z

Palincsar, A. S. (1998). Social Constructivist Perspectives on Teaching and Learning. \textit{Annual Review of Psychology}, \textit{49}(1), 345–375. https://doi.org/10.1146/annurev.psych.49.1.345 

Panadero, E. (2017). A Review of Self-regulated Learning: Six Models and Four Directions for Research. \textit{Frontiers in Psychology}, \textit{8}. https://www.frontiersin.org/articles/10.3389/fpsyg.2017.00422 

Pedaste, M., \& Leijen, Ä. (2018). How Can Advanced Technologies Support the Contemporary Learning Approach? \textit{2018 IEEE 18th International Conference on Advanced Learning Technologies (ICALT)}, 21–23. https://doi.org/10.1109/ICALT.2018.00011 

Pelánek, R. (2017). Bayesian knowledge tracing, logistic models, and beyond: An overview of learner modeling techniques. \textit{User Modeling and User-Adapted Interaction}, \textit{27}(3), 313–350. https://doi.org/10.1007/s11257-017-9193-2 

Pelletier, C. (2024). Against personalised learning. \textit{International Journal of Artificial Intelligence in Education}, 34(1), 111-115. https://doi.org/10.1007/s40593-023-00348-z

Peteros, E. D., Lucino, R. V. A., Yagong, J. H., Bacus, M. J. D., de Vera, J. V., Alcantara, G. A., \& Fulgencio, M. D. (2022). \textit{SELF-REGULATION, SELF-EFFICACY AND STUDENTS’ MATH PERFORMANCE IN MODULAR DISTANCE LEARNING DURING THE COVID-19 PANDEMIC}. \textit{04}(03).

Philpott, A., \& Son, J.-B. (2022). Quest-based learning and motivation in an EFL context. \textit{Computer Assisted Language Learning}, \textit{0}(0), 1–25. https://doi.org/10.1080/09588221.2022.2033790 

Potier Watkins, C., \& Dehaene, S. (2023). Can a Tablet Game That Boosts Kindergarten Phonics Advance 1st Grade Reading? \textit{The Journal of Experimental Education}, 1–24. https://doi.org/10.1080/00220973.2023.2173129 

Puzziferro, M. (2008). Online Technologies Self-Efficacy and Self-Regulated Learning as Predictors of Final Grade and Satisfaction in College-Level Online Courses. \textit{American Journal of Distance Education}, \textit{22}(2), 72–89. https://doi.org/10.1080/08923640802039024

Reeves, T. C., \& Lin, L. (2020). The research we have is not the research we need. \textit{Educational Technology Research and Development}, \textit{68}(4), 1991–2001. https://doi.org/10.1007/s11423-020-09811-3 

Roll, I., Aleven, V., McLaren, B. M., \& Koedinger, K. R. (2007). Designing for metacognition—Applying cognitive tutor principles to the tutoring of help seeking. \textit{Metacognition and Learning}, \textit{2}(2), 125–140. https://doi.org/10.1007/s11409-007-9010-0

Roschelle, J., Feng, M., Murphy, R. F., \& Mason, C. A. (2016). Online Mathematics Homework Increases Student Achievement. \textit{AERA Open}, 2(4), 2332858416673968. https://doi.org/10.1177/2332858416673968

Rosé, C. P., \& Ferschke, O. (2016). Technology Support for Discussion Based Learning: From Computer Supported Collaborative Learning to the Future of Massive Open Online Courses. \textit{International Journal of Artificial Intelligence in Education}, \textit{26}(2), 660–678. https://doi.org/10.1007/s40593-016-0107-y

Ryan, R. M., \& Deci, E. L. (2000). Intrinsic and Extrinsic Motivations: Classic Definitions and New Directions. \textit{Contemporary Educational Psychology}, \textit{25}(1), 54–67. https://doi.org/10.1006/ceps.1999.1020 

Ryan, R. M., \& Deci, E. L. (2020). Intrinsic and extrinsic motivation from a self-determination theory perspective: Definitions, theory, practices, and future directions. \textit{Contemporary Educational Psychology}, \textit{61}, 101860. https://doi.org/10.1016/j.cedpsych.2020.101860 

Sailer, M., \& Homner, L. (2020). The Gamification of Learning: A Meta-analysis. \textit{Educational Psychology Review}, \textit{32}(1), 77–112. https://doi.org/10.1007/s10648-019-09498-w 

Saks, K., \& Leijen, Ä. (2014). Distinguishing Self-directed and Self-regulated Learning and Measuring them in the E-learning Context. \textit{Procedia - Social and Behavioral Sciences}, \textit{112}, 190–198. https://doi.org/10.1016/j.sbspro.2014.01.1155

Salvin, R. E., \& Karweit, N. L. (1985). Effects of Whole Class, Ability Grouped, and Individualized Instruction on Mathematics Achievement. \textit{American Educational Research Journal}, \textit{22}(3), 351–367. https://doi.org/10.3102/00028312022003351 

Seaborn, K., \& Fels, D. I. (2015). Gamification in theory and action: A survey. \textit{International Journal of Human-Computer Studies}, \textit{74}, 14–31. https://doi.org/10.1016/j.ijhcs.2014.09.006 

Shemshack, A., Kinshuk, \& Spector, J. M. (2021). A comprehensive analysis of personalized learning components. \textit{Journal of Computers in Education}, \textit{8}(4), 485–503. https://doi.org/10.1007/s40692-021-00188-7 

Shemshack, A., \& Spector, J. M. (2020). A systematic literature review of personalized learning terms. \textit{Smart Learning Environments}, \textit{7}(1), 33. https://doi.org/10.1186/s40561-020-00140-9 

Shute, V., \& Towle, B. (2003). Adaptive E-Learning. \textit{Educational Psychologist}, \textit{38}(2), 105–114. https://doi.org/10.1207/S15326985EP3802\_5

Smallwood, R. D. (1962). \textit{A decision structure for teaching machines}. 

Soderstrom, N. C., \& Bjork, R. A. (2015). Learning Versus Performance: An Integrative Review. \textit{Perspectives on Psychological Science}, \textit{10}(2), 176–199. https://doi.org/10.1177/1745691615569000 

Srinivasan, V. (2022). AI \& learning: A preferred future. \textit{Computers and Education: Artificial Intelligence}, \textit{3}, 100062. https://doi.org/10.1016/j.caeai.2022.100062 

Stanley, \& Lehman. (2015). \textit{Why Greatness Cannot Be Planned | SpringerLink}. https://link.springer.com/book/10.1007/978-3-319-15524-1 

Steenbergen-Hu, S., \& Cooper, H. (2014). A meta-analysis of the effectiveness of intelligent tutoring systems on college students’ academic learning. \textit{Journal of Educational Psychology}, \textit{106}, 331–347. https://doi.org/10.1037/a0034752 

St-Hilaire, F., Vu, D. D., Frau, A., Burns, N., Faraji, F., Potochny, J., Robert, S., Roussel, A., Zheng, S., Glazier, T., Romano, J. V., Belfer, R., Shayan, M., Smofsky, A., Delarosbil, T., Ahn, S., Eden-Walker, S., Sony, K., Ching, A. O., … Kochmar, E. (2022). A New Era: Intelligent Tutoring Systems Will Transform Online Learning for Millions. \textit{ArXiv:2203.03724 [Cs]}. http://arxiv.org/abs/2203.03724 

Sutarni, N., Ramdhany, M. A., Hufad, A., \& Kurniawan, E. (2021). Self-regulated learning and digital learning environment: Its’ effect on academic achievement during the pandemic. \textit{Jurnal Cakrawala Pendidikan}, \textit{40}(2), 374–388. https://doi.org/10.21831/cp.v40i2.40718 

Tan, S. C., Lee, A. V. Y., \& Lee, M. (2022). A systematic review of artificial intelligence techniques for collaborative learning over the past two decades. \textit{Computers and Education: Artificial Intelligence}, \textit{3}, 100097. https://doi.org/10.1016/j.caeai.2022.100097

Tetzlaff, L., Schmiedek, F., \& Brod, G. (2021). Developing Personalized Education: A Dynamic Framework. \textit{Educational Psychology Review}, \textit{33}(3), 863–882. https://doi.org/10.1007/s10648-020-09570-w 

Theobald, M. (2021). Self-regulated learning training programs enhance university students’ academic performance, self-regulated learning strategies, and motivation: A meta-analysis. \textit{Contemporary Educational Psychology}, \textit{66}, 101976. https://doi.org/10.1016/j.cedpsych.2021.101976 

Tolman, E. C., \& Honzik, C. H. (1930). Introduction and removal of reward, and maze performance in rats. \textit{University of California Publications in Psychology}, \textit{4}, 257–275. 

Tuomi, I. (2023). Beyond Mastery: Toward a Broader Understanding of AI in Education. \textit{International Journal of Artificial Intelligence in Education}. https://doi.org/10.1007/s40593-023-00343-4

Turner, M. (2014). \textit{The Origin of Ideas: Blending, Creativity, and the Human Spark}. Oxford University Press. 

UNESCO. (2017). \textit{Education for Sustainable Development Goals: Learning objectives—UNESCO Digital Library}. https://unesdoc.unesco.org/ark:/48223/pf0000247444 

UNESCO. (2022). \textit{Reimagining education: The international science and evidence based education assessment}. https://unesdoc.unesco.org/ark:/48223/pf0000380985/PDF/380985eng.pdf.multi 

Urh, M., Vukovic, G., Jereb, E., \& Pintar, R. (2015). The Model for Introduction of Gamification into E-learning in Higher Education. \textit{Procedia - Social and Behavioral Sciences}, \textit{197}, 388–397. https://doi.org/10.1016/j.sbspro.2015.07.154 

US Department of Education. (2010). \textit{Transforming American education: Learning powered by technology | VOCEDplus, the international tertiary education and research database}. https://www.voced.edu.au/content/ngv:46439 

US Department of Education. (2017). \textit{Reimagining the role of technology in higher education: A supplement to the national education technology plan}. US Department of Education, Office of Educational Technology. https://tech.ed.gov/netp/ 

van Alten, D. C. D., Phielix, C., Janssen, J., \& Kester, L. (2020). Self-regulated learning support in flipped learning videos enhances learning outcomes. \textit{Computers \& Education}, \textit{158}, 104000. \href{https://doi.org/10.1016/j.compedu.2020.104000}{https://doi.org/10.1016/j.compedu.2020.104000} 

Vandewaetere, M., \& Clarebout, G. (2014). Advanced Technologies for Personalized Learning, Instruction, and Performance. In J. M. Spector, M. D. Merrill, J. Elen, \& M. J. Bishop (Eds.), \textit{Handbook of Research on Educational Communications and Technology} (pp. 425–437). Springer. \href{https://doi.org/10.1007/978-1-4614-3185-5\_34}{https://doi.org/10.1007/978-1-4614-3185-5\_34}

VanLehn, K. (2006). The Behavior of Tutoring Systems. \textit{International Journal of Artificial Intelligence in Education}, \textit{16}(3), 227–265.

VanLehn, K. (2011). The Relative Effectiveness of Human Tutoring, Intelligent Tutoring Systems, and Other Tutoring Systems. \textit{Educational Psychologist}, \textit{46}(4), 197–221. \href{https://doi.org/10.1080/00461520.2011.611369}{https://doi.org/10.1080/00461520.2011.611369}

VanLehn, K. (2011). The Relative Effectiveness of Human Tutoring, Intelligent Tutoring Systems, and Other Tutoring Systems. \textit{Educational Psychologist}, \textit{46}(4), 197–221. \href{https://doi.org/10.1080/00461520.2011.611369}{https://doi.org/10.1080/00461520.2011.611369} 

Vygotsky, L. S., \& Cole, M. (1978). \textit{Mind in Society: Development of Higher Psychological Processes}. Harvard University Press. 

Walkington, C., \& Bernacki, M. L. (2020). Appraising research on personalized learning: Definitions, theoretical alignment, advancements, and future directions. \textit{Journal of Research on Technology in Education}, \textit{52}(3), 235–252. \href{https://doi.org/10.1080/15391523.2020.1747757}{https://doi.org/10.1080/15391523.2020.1747757} 

Wambsganss, T., Kueng, T., Soellner, M., \& Leimeister, J. M. (2021). ArgueTutor: An Adaptive Dialog-Based Learning System for Argumentation Skills. \textit{Proceedings of the 2021 CHI Conference on Human Factors in Computing Systems}, 1–13. \href{https://doi.org/10.1145/3411764.3445781}{https://doi.org/10.1145/3411764.3445781}

Wang, K. D., Wu, Z., Tufts II, L. N., Wieman, C., Salehi, S., \& Haber, N. (2024). Scaffold or Crutch? Examining College Students' Use and Views of Generative AI Tools for STEM Education. \textit{arXiv preprint arXiv:2412.02653}. \href{https://doi.org/10.48550/arXiv.2412.02653}{https://doi.org/10.48550/arXiv.2412.02653}

Watters, A. (2023). \textit{Teaching Machines: The History of Personalized Learning}.

Winne, P. H. (2017). Learning Analytics for Self-Regulated Learning. In \textit{Handbook of Learning Analytics} (pp. 241–249). Society for Learning Analytics Research. 10.18608/hla17.021 

Winne, P. H. (2021). Open Learner Models Working in Symbiosis With Self-Regulating Learners: A Research Agenda. \textit{International Journal of Artificial Intelligence in Education}, 31(3), 446–459. https://doi.org/10.1007/s40593-020-00212-4

Winter, J. w. (2018). Analysis of knowledge construction during group space activities in a flipped learning course. \textit{Journal of Computer Assisted Learning}, \textit{34}(6), 720–730. https://doi.org/10.1111/jcal.12279 

World Economic Forum. (2021). \textit{Building a Common Language for Skills at Work A Global Taxonomy}. 

Wu, T.-T., Lee, H.-Y., Li, P.-H., Huang, C.-N., \& Huang, Y.-M. (2023). Promoting Self-Regulation Progress and Knowledge Construction in Blended Learning via ChatGPT-Based Learning Aid. \textit{Journal of Educational Computing Research}, 07356331231191125. \href{https://doi.org/10.1177/07356331231191125}{https://doi.org/10.1177/07356331231191125}

Xie, H., Chu, H.-C., Hwang, G.-J., \& Wang, C.-C. (2019). Trends and development in technology-enhanced adaptive/personalized learning: A systematic review of journal publications from 2007 to 2017. \textit{Computers \& Education}, \textit{140}, 103599. https://doi.org/10.1016/j.compedu.2019.103599

Zawacki-Richter, O., Marín, V. I., Bond, M., \& Gouverneur, F. (2019). Systematic review of research on artificial intelligence applications in higher education – where are the educators? \textit{International Journal of Educational Technology in Higher Education}, \textit{16}(1), 39. https://doi.org/10.1186/s41239-019-0171-0 

Zhang, K., \& Aslan, A. B. (2021). AI technologies for education: Recent research \& future directions. \textit{Computers and Education: Artificial Intelligence}, \textit{2}, 100025. https://doi.org/10.1016/j.caeai.2021.100025 

Zimmerman, B. J. (1986). Becoming a self-regulated learner: Which are the key subprocesses? \textit{Contemporary Educational Psychology}, \textit{11}(4), 307–313. https://doi.org/10.1016/0361-476X(86)90027-5 

Zimmerman, B. J. (2000). Chapter 2 - Attaining Self-Regulation: A Social Cognitive Perspective. In M. Boekaerts, P. R. Pintrich, \& M. Zeidner (Eds.), \textit{Handbook of Self-Regulation} (pp. 13–39). Academic Press. https://doi.org/10.1016/B978-012109890-2/50031-7 

\end{document}